\documentclass[twocolumn,prb,showpacs]{revtex4}
\usepackage{amssymb,amsmath}
\usepackage{graphicx,float}
\pdfoutput=1
\usepackage{bm}
\usepackage{array}
\usepackage{algorithm}
\usepackage{algorithmic}

\begin{document}


\title{\bf Hidden geometries in networks arising  from cooperative
  self-assembly}
\author{Milovan \v{S}uvakov$^{1,2}$ ,  Miroslav Andjelkovi\'c$^{1,3}$,
  Bosiljka Tadi\'c$^{1}$}
\vspace*{3mm}
\affiliation{$^1$Department of Theoretical Physics, Jo\v zef Stefan Institute,
Jamova 39, Ljubljana, Slovenia }
\affiliation{$^2$Institute of Physics,  University of Belgrade, Pregrevica 118, 11080
  Zemun-Belgrade, Serbia}
\affiliation{$^3${Institute of Nuclear Sciences Vin\v ca,  University
    of Belgrade,  1100 Belgrade, Serbia}}\vspace*{3mm}

\date{\today}

\begin{abstract}
\noindent Multilevel self-assembly involving small structured groups of nano-particles provides new routes to development of functional materials with a sophisticated architecture.  Apart from the inter-particle forces, the geometrical  shapes and compatibility of the building blocks are decisive factors in each phase of growth. Therefore, a comprehensive understanding of these processes is essential for the design of large assemblies of desired properties.  
Here, we introduce a computational model for  cooperative self-assembly with  simultaneous attachment of structured groups of particles, which can be described by simplexes (connected pairs, triangles, tetrahedrons and higher order cliques) to a growing network, starting from a small seed. 
The model incorporates geometric rules that provide suitable nesting spaces for the new group and the chemical affinity $\nu$ of the system to acceping an excess number of particles.
For varying chemical affinity, we grow different classes of assemblies  by binding the cliques of distributed sizes. Furthermore, to characterise the emergent large-scale structures, we use the metrics of  graph theory and algebraic topology of graphs, and 4-point test for the intrinsic hyperbolicity of the networks.
Our results show that higher Q-connectedness of  the appearing simplicial complexes can arise due to only geometrical factors, i.e., for $\nu=0$, and that it can be effectively modulated by changing the chemical potential and the  polydispersity of the size of binding simplexes. 
For certain parameters in the model we obtain networks of mono-dispersed clicks, triangles and tetrahedrons, which represent the geometrical descriptors that are relevant in quantum physics and  frequently occurring chemical clusters.
\\[3pt]
\end{abstract}
\maketitle
\section{Introduction\label{sec-intro}}
Self-assembly of nanoscale objects has been recognised as a powerful
method enabling the design of advanced materials with new 
optical, magnetic, conducting and other properties 
\cite{SA_NANOfocus2012,Sa_allscales_Science2002,SA_colloids_scales2016,SA_hierarchicalMagnPlasmonic_ACSN2011}. Complex materials with a new functionality often exhibit hierarchical architecture \cite{SA_account2012,SA_forcesSmall2015,SA_Flowers,SA_replicationInfoPatterns_Nat2011},
suggesting that the self-assembly occurs at different scales from
individual nanoparticles to groups and clusters to
macroscale materials. In this regard, a cooperative  binding of
small formatted nanoparticle structures is crucial for the
developing large-scale aggregates. They can be prefabricated nanocrystals, self-replicated information-bearing patterns \cite{SA_crystalsSoftMatt2011,SA_collectiveNR_SoftMatt2013,SA_replicationInfoPatterns_Nat2011}, or spontaneously formed groups of nanoparticles
\cite{SA_NANOfocus2012,SA_colloids_scales2016,SA_account2012,SA_collectiveMagnetic_AdvMat2016}. 
The affinity of nanoparticles to merge into a small formation, which then appears as a building block on a larger scale, depends on the particle density and constraints applied in the manufacturing process, and other factors that influence the interactions between them
\cite{SA_forcesSmall2015,SA_Flowers,minEclusters_PCCP2016}.  In addition to binding energy,  this process is regulated  by pertinent geometric rules
\cite{AT_chemistry,Geometry_colloidsPRE2015,AT_Materials_Jap2016}. Therefore, the control of the impact of self-assembly at various levels on the emerging hierarchical structure is essential 
for the new functionality of macroscopic  materials. Here, we use numerical modeling to deepen the understanding of cooperative processes of self-assembly and geometric properties of structures that can arise.

In this context, a suitable presentation by graphs or nanonetworks \cite{we_nanonetworks2013} enables the use of advanced graph theory methods to elucidate the structure and abstract essential geometrical descriptors of nano-structured materials \cite{AT_Materials_Jap2016}. For instance, the network model and topology analysis have proved useful in revealing the structural elements that are responsible for the improved tunneling conduction in self-assembled nanoparticle films \cite{we_NL2007,we_topReview2010,we_Qnets2016}, and to identify hidden order in amorphous materials \cite{Jap2_icosahedrons}.  Some recent investigations show how the use of topology can open new ways for designing materials inspired by mathematics \cite{AT_Materials_Jap2016,AT_Jap_book2015}. 
On the other hand,  the research of growing complex networks has recently been extended to explore the attachment of objects (loops, simplexes) under geometric rules and control parameters
\cite{we_SAloops,SA_modelDNAtiling2016,Geometry_BianconiSR2017,Geometry_KrioukovPRE2010}. 
In this regard, the self-assembly can be understood as a language that can describe the complex architecture of these networks.
Varying the assembly rules and  parameters enables us to explore a broad range of structures, compared to the laboratory experiments and the potential limits of the aggregation process, and understand the emergence of new features
\cite{we_SFloops2006,SA_modelDNAtiling2016,Geometry_BianconiSR2017,Geometry_KrioukovPRE2010}.
A particular anisotropy of the interaction and spatial constraints can lead to some interesting low-dimensional assemblies, for example, chains \cite{SA_chains} and patterns obtained by tiling
or recognition-binding on a two-dimensional lattice \cite{SA_modelDNAtiling2016}, and self-assembly of loops under the planar graph rules \cite{we_SAloops}.
By contrast, self-assembly of geometric objects without spatial embedding can lead to complex, hierarchically organised networks. 
To explore the hidden topology of these networks beyond the standard graph-theory metric \cite{bb-book,SD-book2}, the advanced techniques of algebraic topology of graphs \cite{kozlov-book,jj-book} are used; the main goal is to find out how different geometric elements (simplexes) are mutually combining to make simplicial complexes.  
  Analysis based on algebraic topology of graphs has been used in some recent studies, for example, to describe the hierarchical organization of social graphs \cite{we_PhysA2015} and the structure of the phase-space manifolds near the jamming transition \cite{we_PRE2015}, as well as to adequately quantify the patterns of inter-brain coordination \cite{we_Brain2016} and logically structured knowledge networks \cite{we_Tags2016}.
Moreover,  in the hidden geometry metric of many complex networks, the closeness of the nodes is expressed by the graph's  generalization of negative curvature or hyperbolicity
\cite{Geometry_KrioukovPRE2010,Hyperbolicity_IEEE2016,Geometry_BianconiSR2017}. It plays a significant role in the network's function. For example, a direct survey of the related graphs revealed the impact of negative curvature on metabolic processes \cite{Geometry_NegC_biol} and traffic on the Internet \cite{Geometry_NegC_traffic}. 

Here we introduce a model for the cooperative self-assembly, in which small, ordered  structures of particles are recognized as \textit{simplexes} or full graphs (cliques) of different size that can attach by nesting in a growing network. The process depends on the size of the group  that is formed by the attachment, and it is directed by two ingredients. These are  \textit{geometric factor}, which refers to the availability of the geometrically appropriate sites where the clique can nest along one of its lower-dimensional faces, and the \textit{chemical factor} associated with the affinity of the system for simultaneous binding an excess number of particles. We notice that for a simplex of a given size  the geometric constraints change systematically when the network grows, whereas the chemical affinity affects the actual binding. 
By exploiting the interplay of these elements, we develop various classes of assemblies represented by graphs, and we investigate their structure using graph-theoretic metrics. We show that these structures possess higher combinatorial connectivity, which can be quantified by algebraic topology measures. 
With a large number of examples we demonstrate how the geometrical element that  plays a key role in the appearance of the higher Q-connectedness can be enhanced or reduced by changing the chemical affinity of the assembly.
We also show that these new structures exhibit a global negative curvature or $\delta$-hyperbolicity. 
Our model is defined for poly-dispersive cliques, whose size
varies according to a given distribution in the range from a connected pair of nodes to 12-clique. 
As a particular case, we consider the aggregation of mono-disperse cliques of a given order.
Below are the details of the model explained in formal language of topology; to pesent the model at work, we provide the Web applet \cite{we-ClNets-applet}.

\section{Results and Discussion\label{sec-results}}
\subsection{Computational model\label{sec-rules}}
\begin{figure*}[htb]
\begin{tabular}{cc}
\resizebox{32pc}{!}{\includegraphics{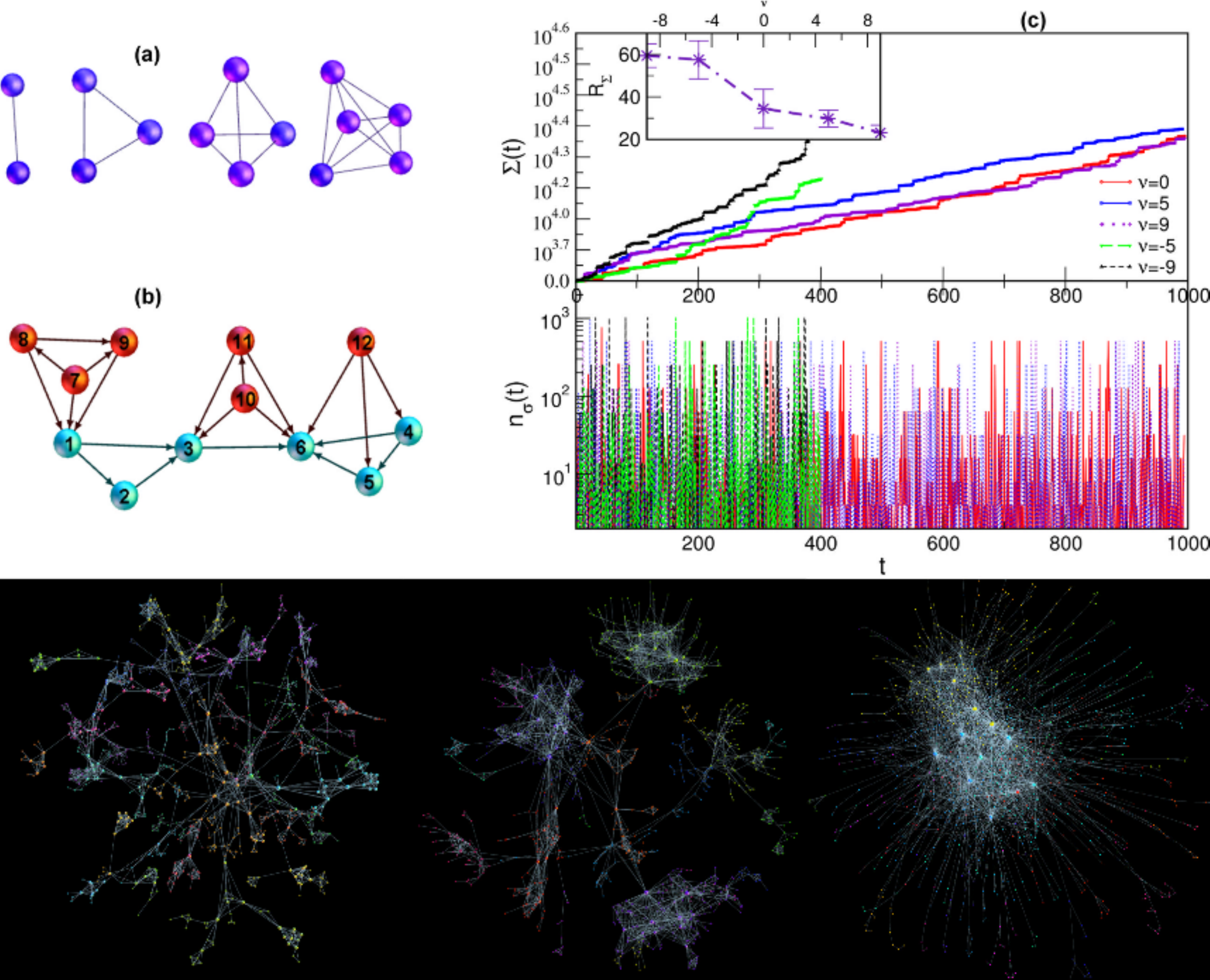}}\\
\end{tabular}
\caption{(a) Examples of geometrical shapes identified as cliques of the order $q_{max}=1,2,3,4$, from left to right. (b) Addition of a
  tetrahedron ($q_{max}=3$) to the system of blue nodes can be nested in three different ways, i.e., by its face of the
  dimension $q=0$, for instance, including the node ``1'',
  $q=1$, the link ``3-6'',   and $q=2$, the triangle ``4-5-6''. The
  corresponding number of new particles $n_a=q_{max}+1-(q+1)=q_{max}-q$ are shown by  red nodes. (c) The number of simplexes $\Sigma (t)$ as function of time for aggregation of poly-disperse cliques at different parameter $\nu$. Lower panel shows the corresponding number $n_\sigma (t)$ of added simplexes per time step. Inset: Average growth rate $R_\Sigma\equiv \langle d\Sigma /dt\rangle$ vs. $\nu$. Bottom
panels:  Networks of aggregated cliques of 
  sizes $s \in [2,10]$ for the varied chemical potential $\nu=
  -9$, $\nu= 0$, and $\nu=+9$, left to right. Different colours of nodes indicate the network's community structure.}
\label{fig-netsing-examples}
\end{figure*}

A clique of order $q_{max}$  is fully connected graph of $s=q_{max}+1$ nodes; some examples are shown in Fig.\ \ref{fig-netsing-examples}. Faces of the clique are cliques of the lower orders which are contained in the original clique $\sigma_q\in \sigma_{q_{max}}$, where $q=0,1,2,3 \cdots q_{max}-1$ correspond to a single node, a pair of connected nodes, a triangle, tetrahedron, etc., up to the largest subgraph of
the order $q_{max}-1$, respectively. The number of equivalent faces is given by $C_q=\binom{q_{max}+1}{q+1}$. Hence, $C_0=q_{max}+1=s$ is the dimension or the number of nodes involved in the considered clique.
We assume that by docking, a clique  \textit{shares a face of order $q$} with an already existing clique in the networks.
In this way, the number of simultaneously added particles (nodes) is given by the difference of the dimension of the clique to be formed by docking and the size of the docking
site, i.e., $n_a=(q_{max}+1)-(q+1)$. Furthermore, we assume that the system's affinity $\nu$ towards adding new particle is finite. Therefore,   the probability of docking along a particular face of the clique is weighted by the factor $e^{-\nu n_a}$, considering the complementary  $n_a$ particles.  Therefore, the normalized probability for docking of a clique of order $q_{max}$ along its face of order $q$ is given by
\begin{equation}
p(q_{max},q;t)= \frac{c_q(t)e^{-\nu (q_{max}-q)}}{\sum _{q=0}^{q_{max}-1}c_q(t)e^{-\nu (q_{max}-q)}} \ ,
\label{eq-pattach}
\end{equation}
where $c_q(t)$ is the number of geometrically similar docking sites of the order $q$ at the evolution time $t$.  
In our model, a clique is formed in each time step $t$; the size of the clique can vary in a given range. In particular, here we
consider cliques of the dimension $s\in [1,12]$ taken from a power-law distribution $g(s)=As^{-2}$, where $A$ is the corresponding normalisation factor.
The empirical fact motivates this form of the distribution, namely, that larger cliques appear less often in modular networks. The network growth by addition of mono-disperse cliques is a particular case of our model.
For instance, by fixing  $s_{min}=s_{max}=3$ (triangles) and $s_{min}=s_{max}=4$ (tetrahedra), we obtain two types of networks with mono-disperse cliques.

The first clique taken from the distribution $g(s)$ is assembled and
considered as the seed structure. Then, at each step, the size of a new clique
$s\in g(s)$ is taken and the clique is formed by attaching the number
$s-q-1$ of new nodes with the selected $q+1$ nodes on the existing
structure.
Then \textit{the  docking condition requires that these $q+1$ nodes match a $q$-face of the
new clique}.
According to Eq.\ (\ref{eq-pattach}), the selection of the simplex of the order $q$ on the current structure depends on the number of geometrically suitable locations and the corresponding weighting factors. Fig.\ \ref{fig-netsing-examples}b illustrates the effects of the geometrical factor in the example of forming a tetrahedron by attachment of $n_a$ red nodes to the small structure shown by the blue nodes.
Considering Eq.\ (\ref{eq-pattach}), the case $\nu=0$ describes the probability of attachment by geometrical factor alone. In this case,  the population of docking sites of the order $q$ determines the likelihood that a new clique will attach by its $q$-face. On the other hand, the number of docking sites of a given size depends on the actual structure of the network.  Note that, by adding a particular clique of the order $q_{max}$ to the system, all its \textit{unshared} faces also appear as new cliques of lower orders. Thus, the number of simplexes fluctuates in time depending not only on the dimension of the clique which is formed in the docking event but also on the size of the actual docking site.
It should be noted that while the simplicial complexes grow through the attachment of new cliques via shared faces, the process can not generate holes and cliques of the order larger than the cut-off size $s_{max}$ of the original distribution $g(s)$. 
In the simulations, we keep track of details constituting  each event. For
example, the small segment of the output file shown below indicates the time step, current network size, the number of simplexes, order of the added clique, the number of new nodes, and list of all nodes
which belong to that clique. 
\begin{verbatim}
24 42 729 2 1 14 42
25 45 785 6 3 28 29 32 43 44 45
26 46 789 3 1 16 28 46
27 47 791 2 1 12 47
28 48 795 3 1 17 36 48
29 49 797 2 1 36 49
30 55 1805 10 6 17 28 30 31 50 51 52 53 54 55
\end{verbatim}
For varied chemical potential $\nu$,  despite the statistically similar population of cliques appearing in the process (taken from the same distribution), the network growth speed and the average rate of the addition of simplexes $R_\Sigma\equiv \langle  d\Sigma (t)/dt\rangle$ are different being dependent on the docking probability. 
Fig.\ \ref{fig-netsing-examples}c displays the evolution of the total number of simplexes $\Sigma (t)$ and the number  $n_\sigma (t)$ of the added simplexes per time step for different networks until they exceed the targeted size of
$N=1000$ nodes for the first time.

Specifically, a fast growth of the network is observed for the negative values of the parameter $\nu$ while much slower growth rates characterize the assembly process for $\nu \geq
0$. In fact, for $\nu <0$ the system ``likes'' addition of new particles, which represent the non-nested parts of the new clique. Hence, the cliques effectively repel each other resulting in a sparse structure and fast growth of the network size and also the addition of new simplexes. In contrast, when  $\nu >0$ the cliques are preferably nested along their larger faces, thus reducing the number of the newly added nodes. This powerful attraction among cliques leads to dense network structure and a small number of added nodes and unshared faces per time step. This situation
results in a slower growth of the network and reduced simplex addition rate, as shown in Fig.\ \ref{fig-netsing-examples}c. In contrast, the case with strictly geometrical assembly, $\nu=0$, 
has no preference for any size of a docking site; the probability is strictly determined by the number of locations of a given size.
 Accordingly, these details of the process have an impact onto the topology of the evolving assemblies, which we study in the following. For illustration, three examples of the networks
containing the number $N\geq 1000$ nodes for varied parameter $\nu$ and the same distribution of the incident cliques are shown in bottom panels in Fig.\ \ref{fig-netsing-examples}.

\subsection{Combinatorial topology of aggregates with poly-disperse cliques\label{sec-topology}}
As the network examples in Fig.\ \ref{fig-netsing-examples} (bottom) demonstrate, the structure
that emerges in the assembly of cliques depends strongly on the
affinity for the simultaneous attachment of many
nodes, apart from the geometrical constraints. 
Specifically, for large negative values of the parameter $\nu$, an active 'repulsion' between the cliques results in the sparse structure, nearly representing a tree of cliques of different orders.  This kind of structures possesses a significant average distance, the modularity, and clustering coefficient, which can be related to the original population of cliques. On the other
hand, for the positive values of $\nu$, the cliques firmly attach to each other, resulting in a gradually smaller number of the simultaneously added particles. The appearing structure possesses a large core of densely packed higher-order cliques while low-order structures remain at the periphery. An impressive network architecture with well-separated communities appears for $\nu=0$, assembled under geometrical
constraints alone. As described below, the graph properties are tunned between these extremes by varying the parameter $\nu$.  

\begin{figure}[!htb]
\begin{tabular}{cc} 
\resizebox{18pc}{!}{\includegraphics{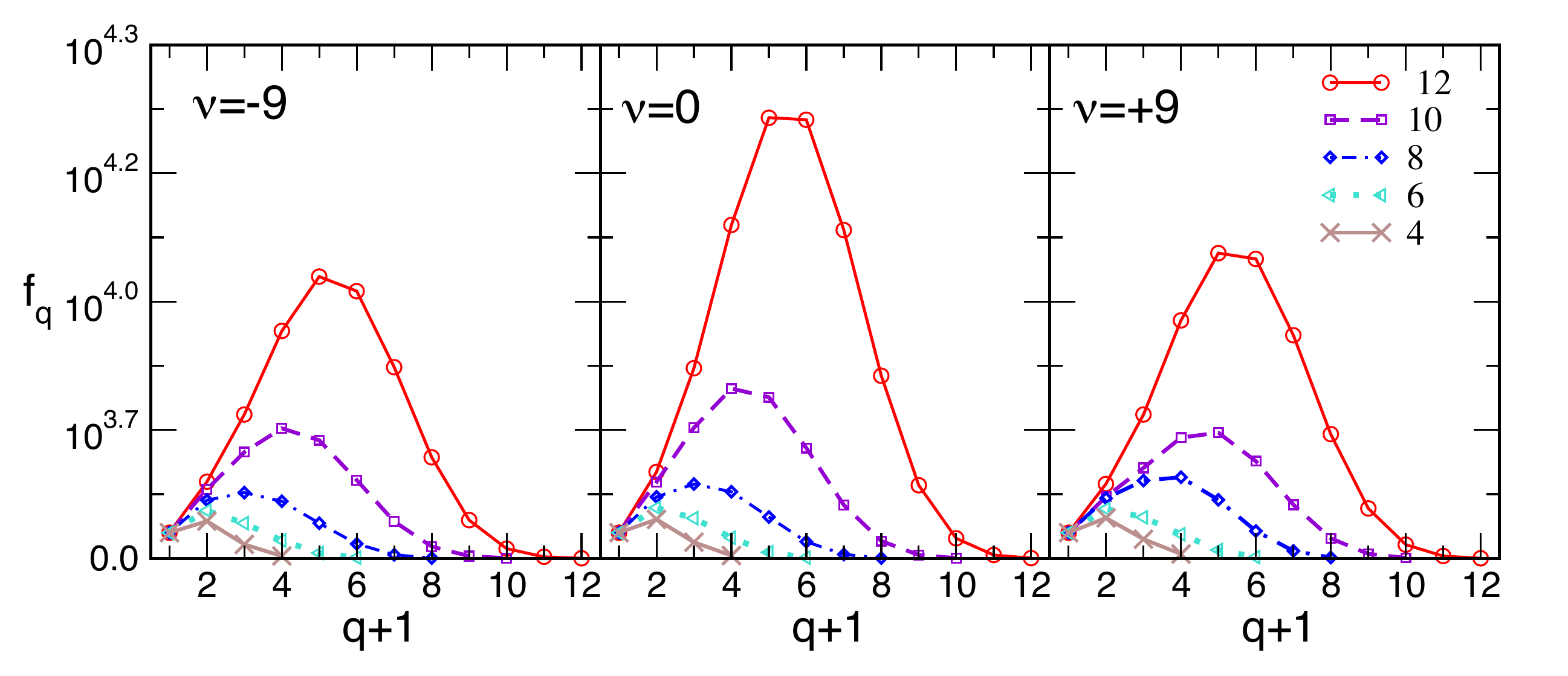}}\\
\end{tabular}
\caption{Topological response function $f_q$ plotted against simplex dimension
  $q+1$ for $\nu =$ -9,  0, and +9; different curves correspond to
  the varied upper dimension of the building cliques $s_{max}$
  indicated in the legend.}
\label{fig-clNets_fq}
\end{figure}

Here, our focus is on the appearance of higher combinatorial topologies of these graphs, which is directly related to the ways that the assembled cliques share their faces of different orders. In the simulations, we keep track of each added
simplex and nodes that participate in it, as explained
above. In this way, for a clique of the order $q_{max}$ we can
distinguish the number of its shared faces of order $q <
q_{max}$. Intuitively, when the groups repel each other, i.e., for $\nu <0$, their common faces will be the lowest orders, such as single nodes and links and, less often, triangles or higher structures.  The opposite situation typically occurs
for $\nu >0$ where the simplexes have a high affinity towards sharing nodes; cf. structure in Fig.\ \ref{fig-netsing-examples}. 
Due to shared faces, for instance, of the order $q$, the number of distinguishable simplexes of that order is smaller than the number of faces $C_q$ of a free added clique. Therefore, the \textit{topological response} function $f_q$ of the network \cite{we_PRE2015} can be determined as the number of different simplexes \textit{at} the topology level $q$; it provides a good measure of the combinatorial complexity of the assembly in response to the varying external parameters $\nu, s_{max}$.  In Fig.\
\ref{fig-clNets_fq}, we show how the function $f_q$ varies along the topology levels $q$ depending on the parameter $\nu$ and the range of the distribution of the attaching cliques. The peak of the distribution shifts towards higher values when larger dimension cliques appear, whereas the height depends on the way that they interconnect at each the topology level.
\begin{figure}[htb]
\begin{tabular}{cc} 
\resizebox{14pc}{!}{\includegraphics{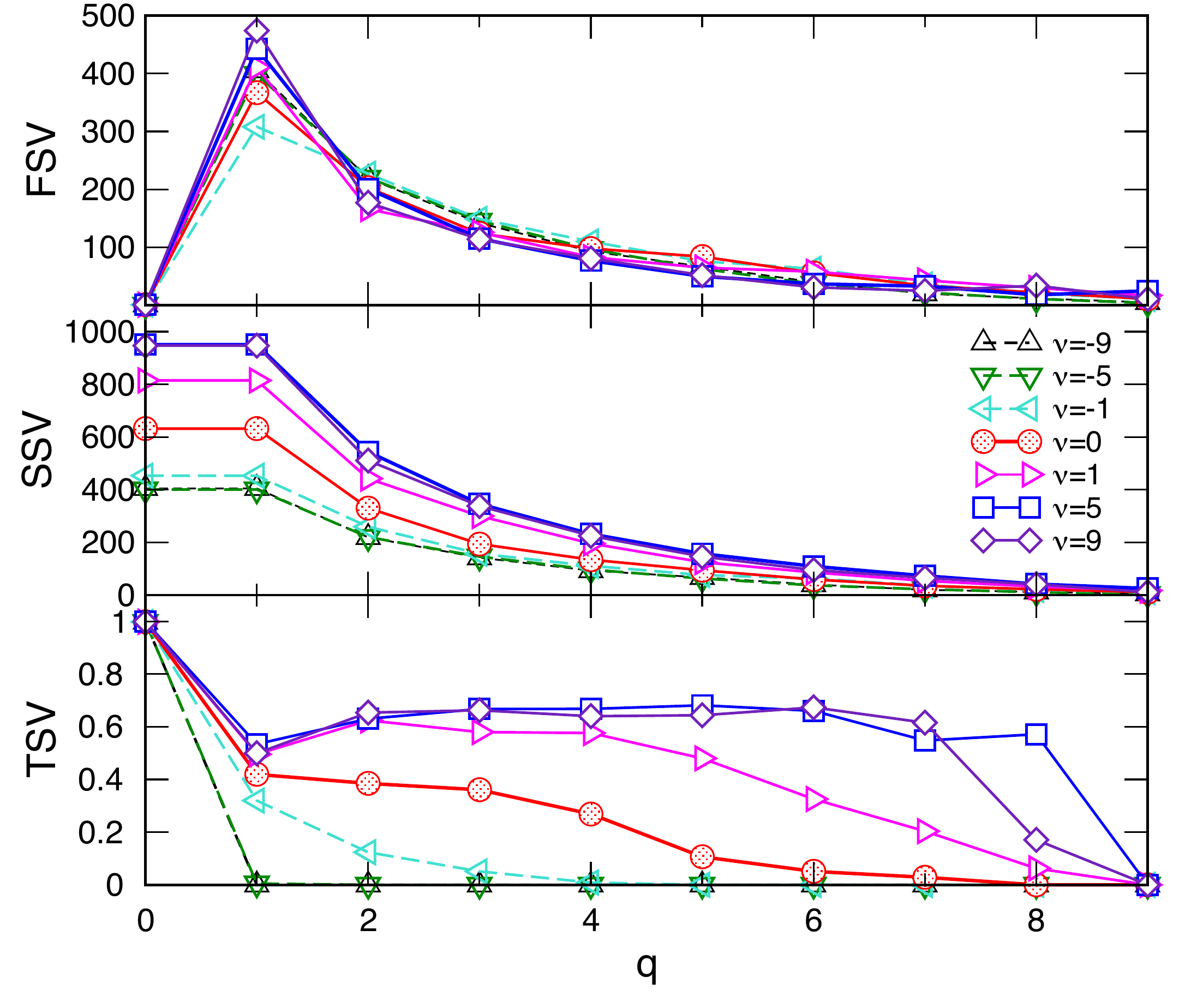}}\\
\end{tabular}
\caption{Components of the first (FSV), second (SSV) and third (TSV) structure vector
  corresponding to the topology level $q$ against $q$  of the networks
  grown at  different values of  $\nu$ and the fixed
  distribution of clique size $s \in [2,10]$.}
\label{fig-clNets_SVs}
\end{figure}
\begin{figure*}[!htb]
\begin{tabular}{cc} 
\resizebox{36pc}{!}{\includegraphics{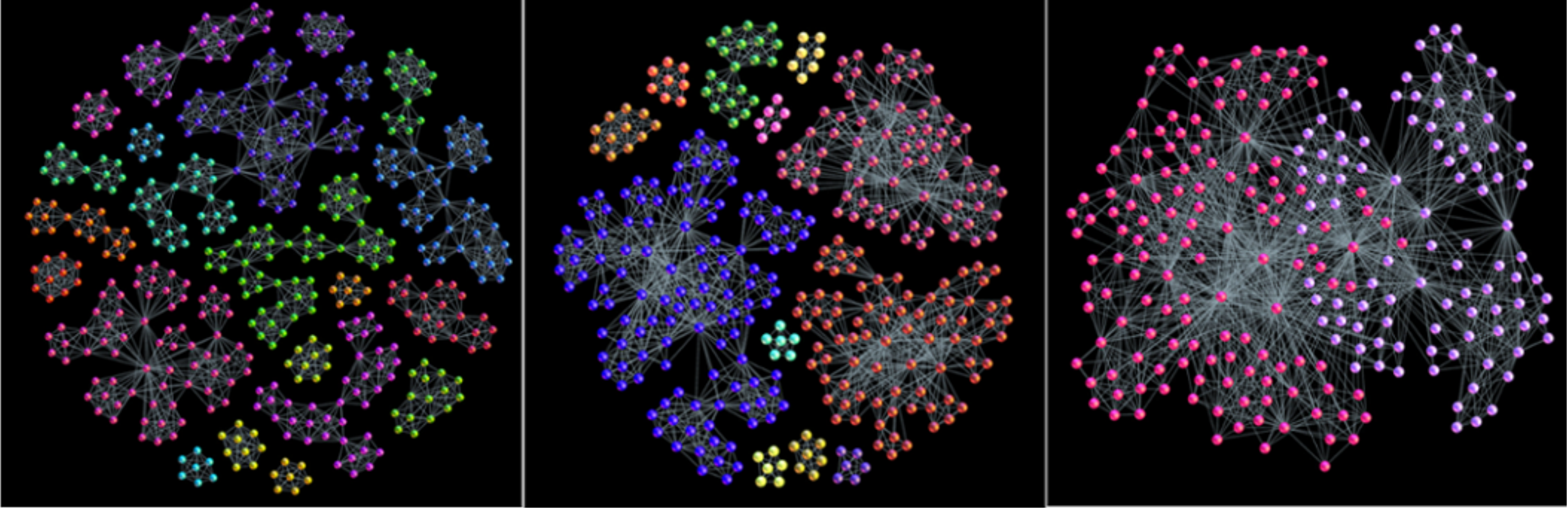}}\\
\end{tabular}
\caption{Adjacency matrix of the network's nodes which participate in
  structures that are ``visible'' at the topology level q=5.
  left to right: $\nu=$-1, 0, and +1. Different colors identify clusters or communities.}
\label{fig-q5cutts}
\end{figure*}
Further, Q-analysis based on the algebraic topology of graphs
\cite{atkin1976,Q-analysis-JJ,Q-analysis-book1982,jj-book} is here applied for characterization of the graph architecture by determination of $q$-connected components for each topology level $q$. 
Specifically, for the topology levels $q=0,1,2,\cdots q_{max}$  of each studied network we determine the components  of three structure vectors, $\{Q_q\}$, $\{n_q\}$  and $\hat{Q_q}$, defined in Methods.
These vectors allow a direct comparison of the hierarchical structure of various emergent networks. 
In Fig.\ \ref{fig-clNets_SVs}, we plot the components of 
these  structure vectors as a function of $q$ for several
assemblies of poly-disperse cliques with different chemical potential $\nu$.

The similarity in the number of $q$-connected components (FSV) reflects the statistically similar population of cliques of all
dimensions (taken from the same distribution) in all studied
networks. However, their inherent structure is significantly
different, which is expressed by the components of SSV and TSV for
various $q$ (see Methods). 
Notably, the third structure vector in networks for $\nu <0$  has
non-zero components only at  lowest topology levels; this implies that different higher-order cliques present in the graph will be separated from each other by removing
the structures of the order $q=1$ (link) between them. The situation is much different in the assemblies grown when $\nu >0$ where the simplicial complexes
containing the higher-order cliques remain
strongly interconnected until the before-last level $q_{max}-1$ =8. 
These findings agree with the impact of the chemical potential favoring the cliques attraction for $\nu >0$ and repulsion for $\nu <0$. In this context, it is interesting to note that structure that was grown solely under the geometrical rules ($\nu=0$) already possesses a sizable hierarchical organization of simplicial complexes; although the degree of connectivity is systematically lower than in the case $\nu =+1$, the structure holds together until the level $q=7$. (See Table\ \ref{tab-SVs-m101} for the exact values). As the Fig.\ \ref{fig-clNets_SVs} shows, this hierarchical architecture of the assembled networks
gradually builds with increasing values of the parameter $\nu$. 
To illustrate the differences in the hierarchical organization of the systems for $\nu=$ -1, 0, +1, in Fig.\ \ref{fig-q5cutts} we display those parts of their structure that are still visible at the topology level $q=5$. Precisely, the nodes participating in the simplexes of order $q\leq 5$ which are not faces of the cliques of the order $q>5$, are removed. The connections among the remaining nodes are shown according to the network's adjacency matrix.

\begin{figure}[htb]
\begin{tabular}{cc} 
\resizebox{18pc}{!}{\includegraphics{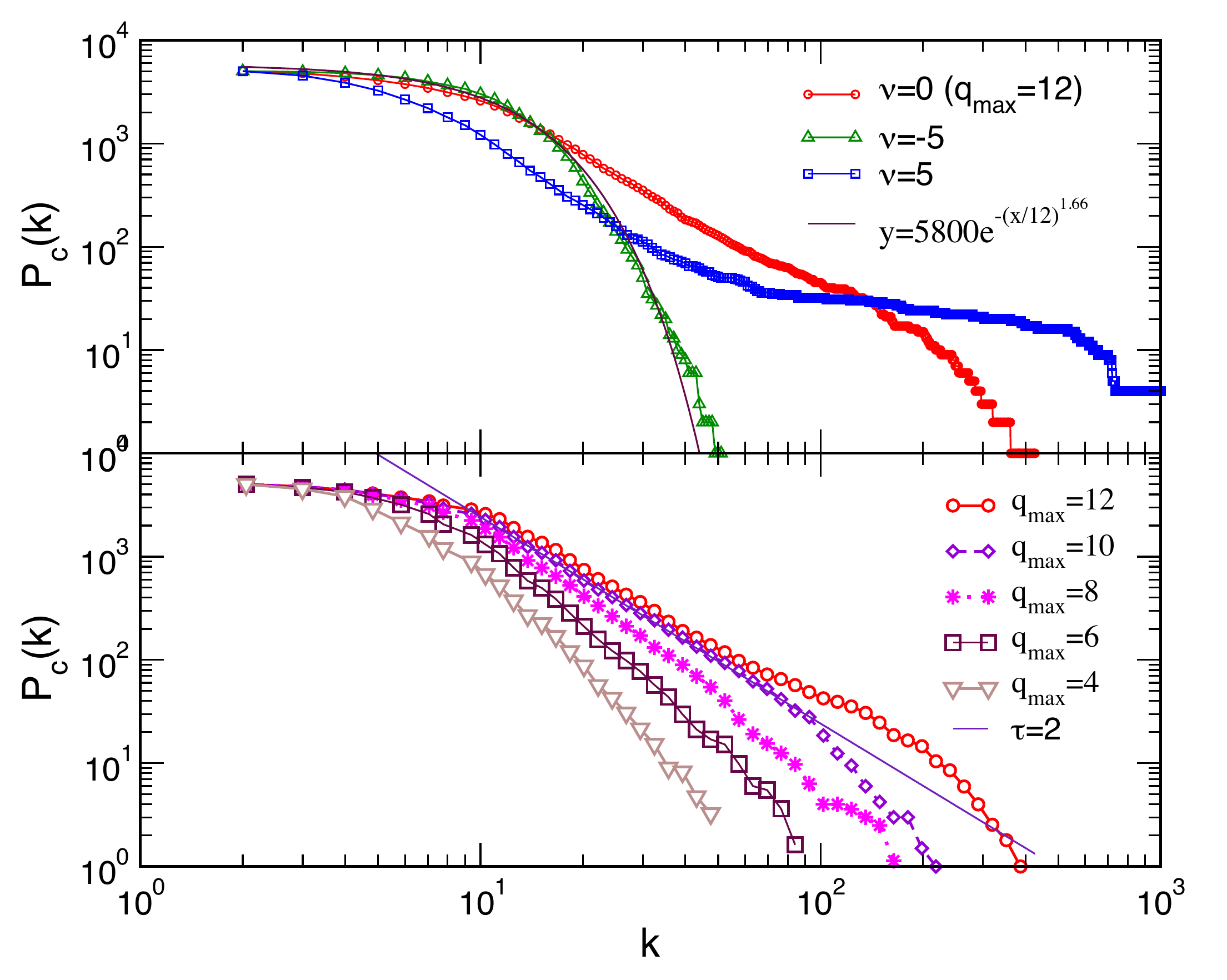}}\\
\end{tabular}
\caption{Cumulative distributions of  the degree in networks of
  aggregated poly-disperse cliques $s \in [2,12]$ and varied chemical
  potential $\nu$ (top panel) and for purely geometrical aggregation
  ($\nu=0$) and varied size of the largest added clique $s_{max}$  (lower panel). Each distribution is averaged over several samples of the networks with the number of nodes $N\geq 5000$.}
\label{fig-clNets_Pkc}
\end{figure}

The node's participation in building various simplexes also manifests in the global statistical features of the network.
The cumulative degree distribution for several studied aggregates is given in Fig.\ \ref{fig-clNets_Pkc}. It is averaged over several realizations of the systems containing over $5000$, where $s_{max}\in [2,12]$. Although a broad distribution of the node's degree occurs in each case, it strongly varies with the parameter $\nu$.
It is interesting to note that, in the networks grown by geometrical constraints with $\nu=0$, we obtain the distribution with a power-law decay $\tau +1\approx 3$ (within the numerical error bars); its cut-off appears to depend on the size of the largest clique. In contrast, the exponential decay is observed for $\nu <0$ while a structure containing many nodes of a large degree is present in the case of clique attraction for $\nu>0$, which is separated from the low-degree nodes. 
Other graph theoretic measures also vary accordingly.

\subsection{$\delta$-Hyperbolicity of the emergent networks\label{sec-hyperbolicity}} 
\begin{figure}[!htb]
\begin{tabular}{cc} 
\resizebox{18pc}{!}{\includegraphics{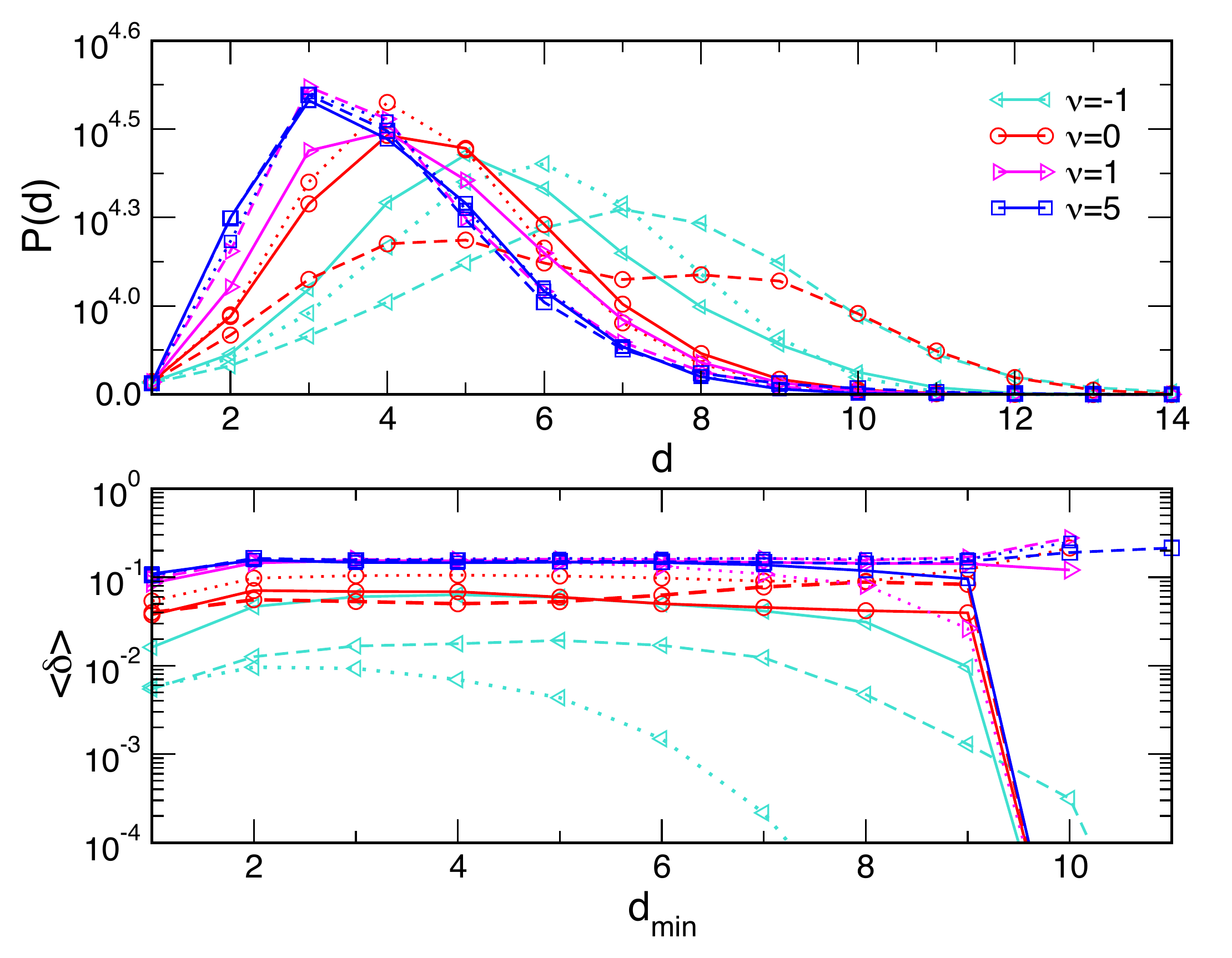}}\\
\end{tabular}
\caption{Histograms of shortest distances $d$ between pairs of nodes
  (top)  and average
  hyperbolicity  $\langle \delta \rangle $ vs. $d_{min}$
   (bottom) in three samples of networks for $\nu=$ -1, 0, +1, and
   +5. Network size is above $N=500$ nodes.}
\label{fig-hyperbolicity-ni}
\end{figure}

For network structures, $\delta$-hyperbolicity is a generalization of negative curvature in the large \cite{Hyperbolicity_IEEE2016}.
Here, we consider the aggregates of cliques, which are known $0$-hyperbolic graphs; therefore, these structures are expected to exhibit this intrinsic property at a larger scale. Following
the procedure described in \cite{Hyperbolicity_IEEE2016},  we investigate the 4-point Gromov hyperbolicity of different emergent networks. Specifically, we determine the average hyperbolicity  $\langle \delta \rangle$ in comparison to the graph's diameter for $\nu=$ -5, -1, 0,+1, and +5, by a sampling of $10^9$ sets of four nodes, as described in Methods. Considering
three different realizations of the network for each $\nu$, we find numerically that  $\delta$ can take the values $\{0,1/2,1\}$; hence, the maximum value $\delta _{max}=1$ suggests that
these assemblies are $1$-hyperbolic. 
In Fig.\
\ref{fig-hyperbolicity-ni} (bottom panel) we plot  the average hyperbolicity $\langle \delta \rangle$ against the minimal
distance  $d_{min}$ of the involved pair in the smallest sum ${\cal{S}}$, see Methods. 
Notably, for all network types $\langle \delta \rangle$  remains bounded at small values.  
In particular, we find that $\langle \delta \rangle = 0$ for the tree graph of cliques corresponding to  $\nu=-5$. Whereas, the hyperbolicity parameter is close to zero in the sparse network of cliques for $\nu =-1$,  and slightly increases in the more compact structures corresponding to $\nu=0$ and $\nu>0$. Note that due to a small number of pairs of nodes having  the largest distance in the graph we observe the fluctuation of 
$\langle \delta \rangle \in [0,0.5]$. The histograms of distances
between all pairs of nodes in the considered networks are also shown in Fig.\ \ref{fig-hyperbolicity-ni} (top panel).

\subsection{Aggregation of monodisperse cliques\label{sec-tetrahedra}}
\begin{figure}[!htb]
\begin{tabular}{cc} 
\resizebox{16pc}{!}{\includegraphics{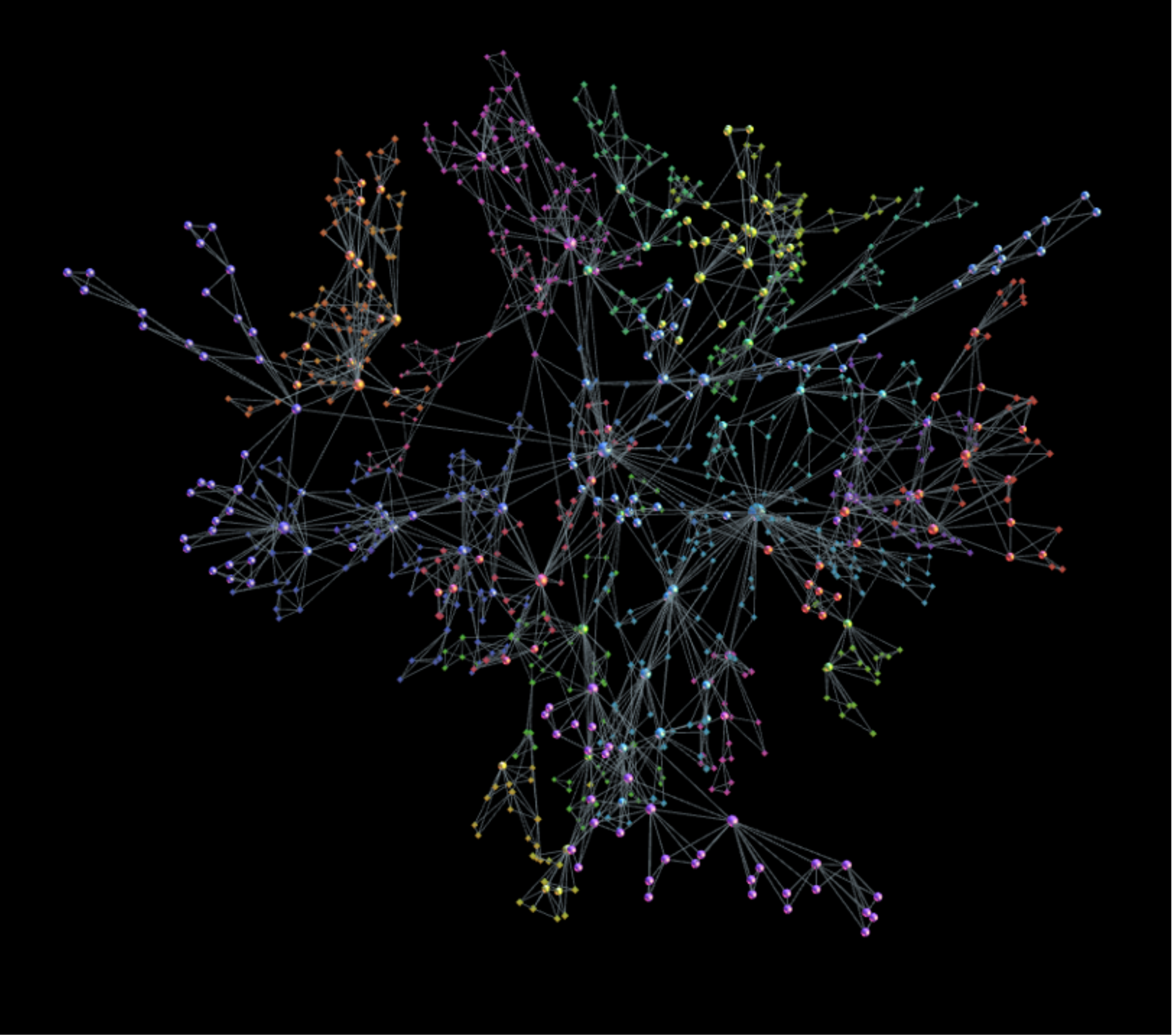}}\\
\resizebox{16pc}{!}{\includegraphics{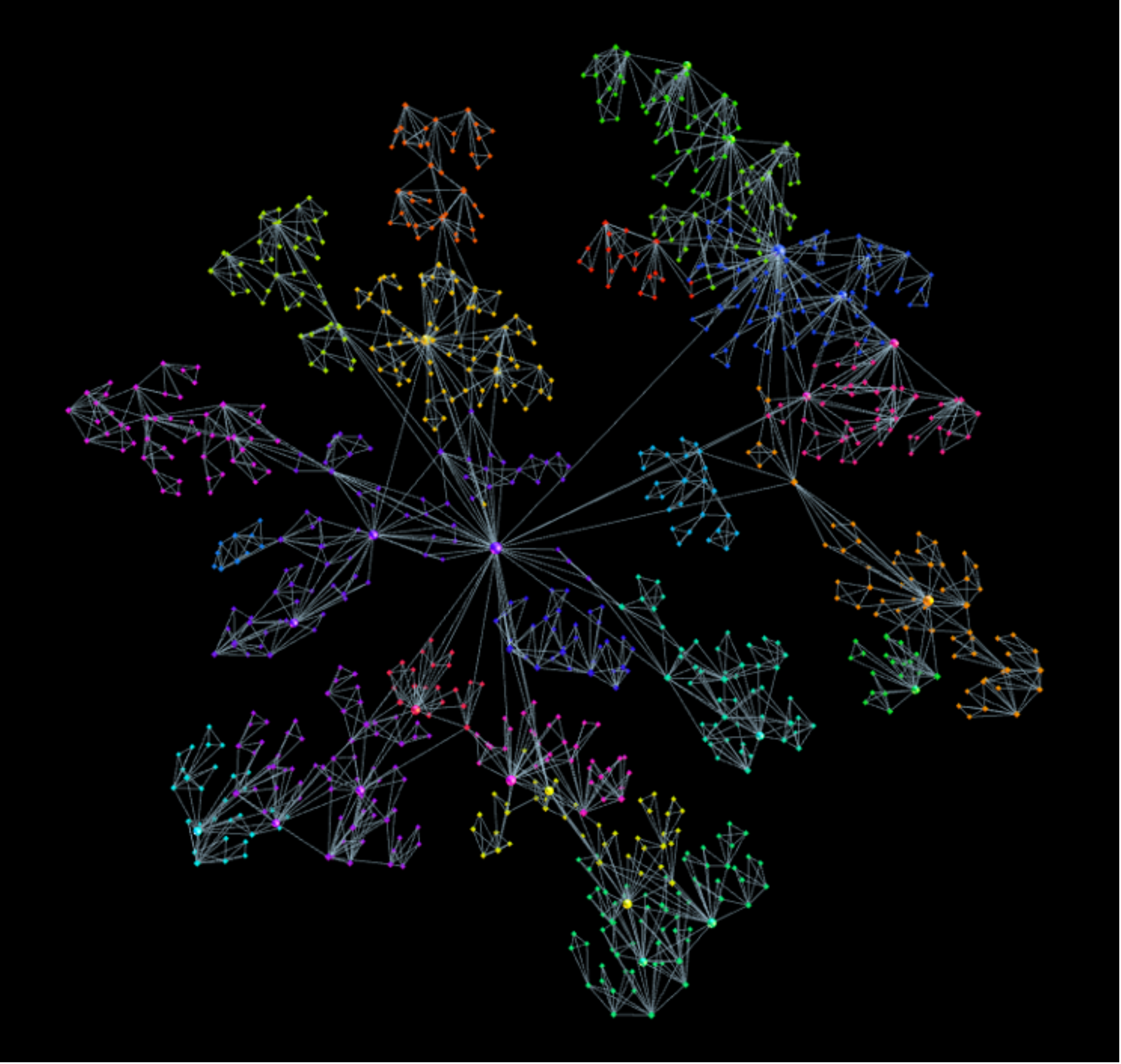}}\\
\end{tabular}
\caption{Aggregates of tetrahedra (top) and triangles (bottom) for $\nu$=0.0.}
\label{fig-monoNets}
\end{figure}

In this section, we briefly consider the structures grown with the same aggregation rules but with mono-disperse building blocks. Some compelling examples are the aggregates of tetrahedra and triangles. Tetrahedral forms are ubiquitous minimum-energy clusters of covalently bonded materials \cite{minEclusters_PCCP2016}.
 We also study the impact of the chemical potential in the event of aggregation of triangles. The importance of triangular geometry was recently pointed  in the context of quantum networks
 \cite{Geometries_QuantumPRE2015}. Some examples of these structures grown by the aggregation rules of our model are shown in Fig.\ \ref{fig-monoNets}. 

\begin{figure}[htb]
\begin{tabular}{cc} 
\resizebox{16pc}{!}{\includegraphics{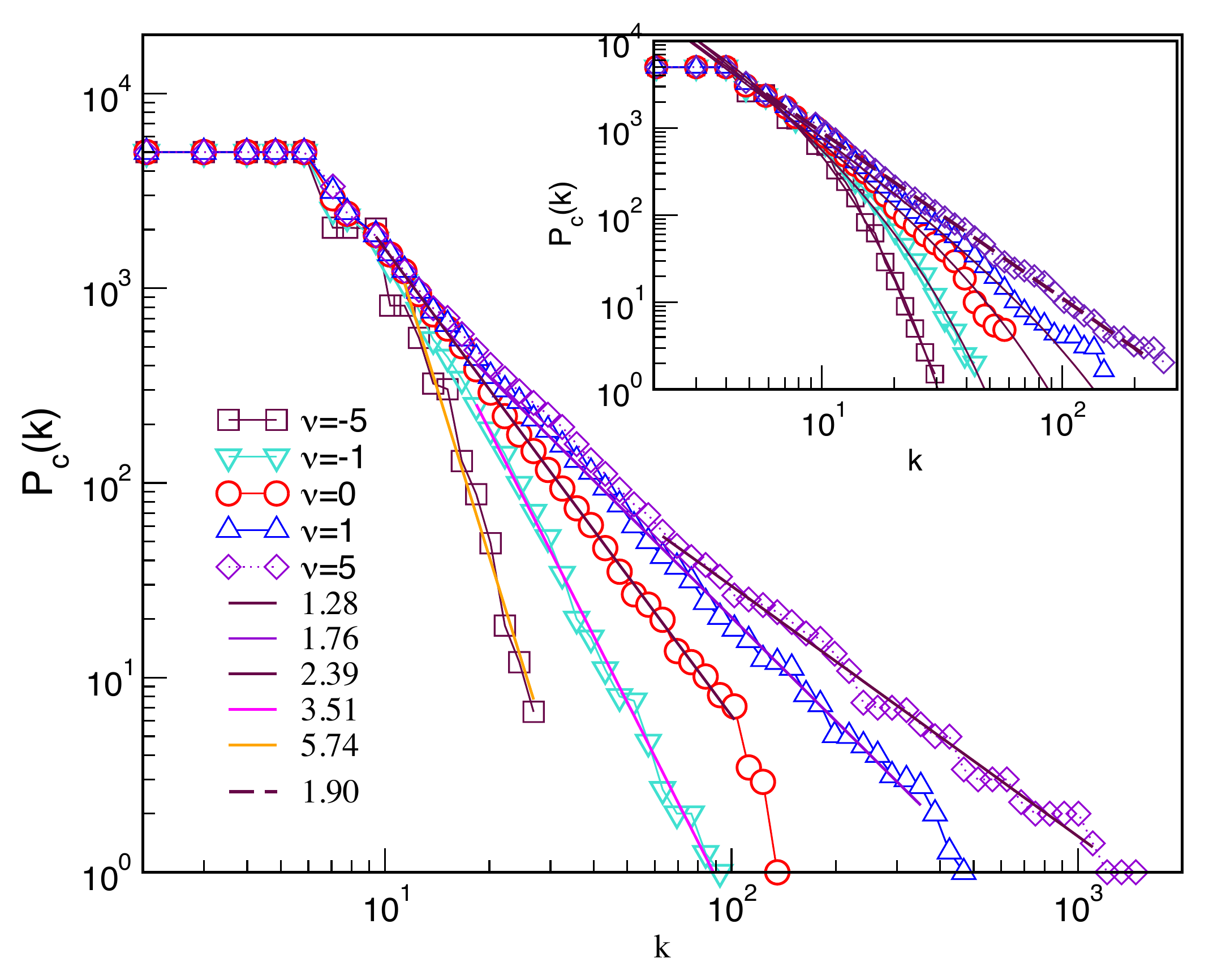}}\\
\end{tabular}
\caption{Cumulative distributions of the node's degree in networks of aggregated mon-disperse cliques (main panel) tetrahedra, and (inset) triangles, for different values of the 
  chemical potential $\nu$. Sample averaging and the number of nodes  $N\geq 5000$ applies. Thick broken and full lines indicate the range where the slopes given in the legend are measured (within the maximal error bars $\pm 0.07$)}
\label{fig-Pkc-mono}
\end{figure}

Since the aggregation process does not alter the size of the largest clique, these networks have only few topology levels. Specifically, in the aggregates of tetrahedra $q_{max}=3$, and they can share nodes, links, and triangles as faces of lower orders; for triangles, $q_{max}=2$ and shared faces are links and nodes. Therefore, their structure vectors are rather short. However, they possess a captivating structure of simplicial complexes, depending on the chemical potential and geometry constraints.
Consequently, the degree distributions are altered by changing $\nu$, as shown in  Fig.\ \ref{fig-Pkc-mono}.  Notably, the appearance of some scale-invariant structures is favored by the mutual attraction of cliques for $\nu >0$. The aggregation of tetrahedra more efficiently builds such structures as compared with triangles. Whereas, the scale-free range is limited with the
exponential cut-offs in the case of triangles unless $\nu$ is
sufficiently large. Further analysis of these and other networks of mono-disperse simplexes is left for future work.

\section{Conclusions\label{sec-conclusions}}

We have introduced a computational model for cooperative self-assembly where small, formed groups of particles appear as building blocks for a large-scale structure. In this context, in addition to the binding forces, the geometric constraints exerted by the rising architecture play an important role  on the proper nesting of the added block. Different geometrically suitable options for nesting a given block structure are further altered due to the chemical affinity of the system for receiving an excess number of particles. 
We have demonstrated how different assemblies with a complex architecture can be formed in the interplay of these geometric and chemical factors.  Moreover, the systematic mapping of the growing structure to the graph not only helps us  formally implement the self-assembly process, but also provides ways to adequately investigate the new structure by means of  advanced graph theory and algebraic topology methods. 
It is interesting that the complex structure of the assembly that possesses combinatorial topology of higher order can arise due to only geometric factors. These topology features are further enhanced in chemically enforced compaction, and, on the contrary, are gradually reduced in sparse networks resulting from chemically favoured repulsion  between building blocks.
Moreover, depending on the dispersion of the components and the chemical factor, the new assemblies may possess scale invariance and an intrinsic global negative curvature; these features are important for their practical use and functionality. Our model with graph-based representation provides a better insight into the mechanisms that drive the assembly of hierarchically organised  networks with higher topological complexity, which is a growing demand for technological applications.

\section{Methods\label{sec-methods}}

\textit{Program flow for clique aggregation} \cite{we-ClNets-applet} 

\begin{algorithm}
\caption{Program Flow: Growth of the graph by attaching simplexes}
\label{alg1}
\begin{algorithmic}[1]
\STATE \textbf{INPUT:} $\nu$, $s_{min}$, $s_{max}$, $N_{max}$
\STATE initialise graph ${\cal{G}}$ as a simplex of size $s$ taken from the distribution $p(s) =As^{-2}$, which is  defined in the range $[s_{min},s_{max}]$
\WHILE {$N<N_{max}$}
\STATE select new simplex size $s_{new}\in [s_{min},s_{max}]$ from the distribution $p(s)$ 
\FORALL {simplexes $\sigma \in {\cal{G}}$ whose order $q_\sigma < s_{new}-1$}
\STATE compute $p_\sigma(s_{new}, q_\sigma)=\exp{-\nu(s_{new}-q_\sigma -1)}$ 
\ENDFOR
\STATE normalise the probability  such that $\sum_{\sigma} p_\sigma(s_{new},q_\sigma)=1$
\STATE select the docking site $\sigma \in p_\sigma(s_{new},q_\sigma)$
\STATE form a new simplex $\sigma_{new}$ by attaching $s_{new}-q_\sigma-1$
new nodes  to the $q_\sigma+1$ nodes of the docking simplex $\sigma$
\STATE sampling the data of interest;
\ENDWHILE
\STATE {\bf END}
\end{algorithmic}
\end{algorithm}

\textit{Q-analysis: definition of structure vectors}\\
To describe the global graph's connectivity at different topology levels $q=0,1,2 \cdots q_{max}$, Q-analysis uses notation from algebraic topology of graphs 
\cite{atkin1976,Q-analysis-JJ,Q-analysis-book1982,jj-book}. 
Specifically, the first structure vector $Q_q$ represents the number of $q$-connected components and the second structure vector  $n_q$ is defined as the number of simplexes of the order greater than or equal to $q$. In this context, two simplexes are $q$-connected if they share a face of the order $q$, i.e., they have at least $q+1$ shared nodes. Then the third structure vector  determined as $\hat{Q_q}\equiv
1-Q_q/n_q$ measures the degree of connectivity at the
topology level $q$ among the higher-order simplexes. 
From the adjacency matrix of a considered graph, we construct incidence matrix by Bron-Kerbosch algorithm \cite{BK}, where simplexes are identified as maximal complete subgraphs (cliques). Then the dimension of the considered simplicial complex equals the dimension of the largest clique $q_{max}+1$ belonging to that complex.

\textit{Measure of curvature: $\delta$-hyperbolicity definition}\\
Following the studies in \cite{Hyperbolicity_IEEE2016} and references there, we implement an algorithm which uses the Gromov's hyperbolicity criterion. Specifically, for an arbitrary set of four nodes A, B, C, and D, the distances (shortest path lengths) between distinct pairs of these nodes are combined in three ways and ordered. For instance,
$$d(A, B) + d(C, D) \le d(A, C) + d(B, D) \le d(A, D) + d(B, C).$$
We denote the largest value ${\cal{L}} = d(A, D) + d(B, C)$, the middle ${\cal{M}} = d(A, C) + d(B, D)$, smallest ${\cal{S}} = d(A, B) + d(C, D)$, and the smallest pair distance of ${\cal{S}}$ as $d_{min}=min\lbrace d(A,B),d(C,D)\rbrace$. Then the  graph is $\delta$-hyperbolic if there is a fixed value $\delta$ for which any four nodes of the graph satisfy the 4-point condition:
\begin{equation}
\frac{{\cal{L}}-{\cal{M}}}{2} \le \delta.
\label{eq-hyperbolicity}
\end{equation}
There is a trivial upper bound $({\cal{L}}-{\cal{M}})/2\le
d_{min}$. Hence, by plotting $({\cal{L}}-{\cal{M}})/2$ against $d_{min}$ we can investigate the worst case growth of the function. 
For a given graph, we first compute the matrix of distances between all pairs of nodes; then, by sampling a large number of sets of nodes for the 4-point condition (\ref{eq-hyperbolicity}) we determine and plot the average $\langle \delta
\rangle$  against  the corresponding distance $d_{min}$.

\textit{Graphs visualisation} We used
 {\texttt{gephi.org}} for graph presentation and community
structure detection by maximum modularity method \cite{blondel2008}.

\section*{\normalsize Acknowledgments}

The authors acknowledge the financial support from the Slovenian
Research Agency under the program P1-0044
and  from the Ministry of Education, Science and Technological Development of
the Republic of Serbia, under the projects OI 171037,  III 41011 and OI 174014.


\appendix
\begin{table}[h]
\begin{tabular}{|l|ccc|ccc|ccc|ccc|ccc|ccc|}
\hline
   & \multicolumn{3}{c|}{\textbf{$\nu=-1$}}                                                  & \multicolumn{3}{c|}{\textbf{$\nu=0$}}                                                  & \multicolumn{3}{c|}{\textbf{$\nu=+1$}}       \\ \hline
q  & \multicolumn{1}{l}{$Q_q$} & \multicolumn{1}{l}{$n_q$} & \multicolumn{1}{l|}{$\hat{Q_q}$}
   & \multicolumn{1}{l}{$Q_q$} & \multicolumn{1}{l}{$n_q$} &\multicolumn{1}{l|}{$\hat{Q_q}$} 
& \multicolumn{1}{l}{$Q_q$} & \multicolumn{1}{l}{$n_q$} &
                                                        \multicolumn{1}{l|}{$\hat{Q_q}$}
  \\ \hline
0& 1&453&0.997&    1&    632&0.998  & 1     &815 &0.999 \\ 
1& 308&453& 0.320&367&632&0.419 &  411   &  815 & 0.495 \\
2& 227&259&0.124& 203 &330  &0.385   & 166     & 442 &0.624  \\
3&149& 157& 0.051& 124 &194  &0.361   &126      & 300  &0.580  \\
4& 110& 111&0.009&  98& 134 & 0.269  & 83     & 196  &  0.576\\
5& 76& 76& 0.000& 84 &94  &0.106   & 65     & 125  & 0.480 \\
6&63& 63& 0.000&56 & 59 &0.051   &  58     &86   &0.325  \\
7& 36& 36& 0.000& 34&35  & 0.029  &  43     & 54  &0.204  \\
8& 20& 20& 0.000&  22&22  &0.000   &  30    &32   &0.063  \\
9& 11& 11& 0.000& 11&11  & 0.000  &  17    & 17  &0.000  \\
\hline
\end{tabular}
\caption{ The components of the  three structure vectors for the
  networks generated at different chemical potential $\nu$.}
\label{tab-SVs-m101}
\end{table}

\end{document}